\documentclass[manuscript,acmsmall]{acmart}
\AtBeginDocument{%
  }

\usepackage{algorithmic}
\usepackage{graphicx}
\usepackage{textcomp}
\usepackage{xcolor}
\usepackage{xspace}
\usepackage{multirow}
\usepackage{subfig}

\setcopyright{acmlicensed}
\copyrightyear{2018}
\acmYear{2018}
\acmDOI{XXXXXXX.XXXXXXX}

\acmJournal{JACM}
\acmVolume{37}
\acmNumber{4}
\acmArticle{111}
\acmMonth{8}

\begin{document}

\title{LogCopilot: Automating Log Aggregation Analysis through Large Language Models}


\author{Senyu Xie}
\affiliation{%
  \institution{Fudan University}
  \city{Shanghai}
  \country{China}}
\email{24210240059@m.fudan.edu.cn}
\orcid{0009-0003-3331-7733}

\author{Chenxi Zhang}
\authornote{Chenxi Zhang is the corresponding author.}
\affiliation{%
  \institution{Xidian University}
  \city{Xi'an}
  \country{China}}
\email{zhangchenxi@xidian.edu.cn}

\author{Tong Zhou}
\affiliation{%
  \institution{Fudan University}
  \city{Shanghai}
  \country{China}}
\email{tongzhou21@m.fudan.edu.cn}

\author{Jiacheng Liu}
\affiliation{%
  \institution{Fudan University}
  \city{Shanghai}
  \country{China}}
\email{25213050272@m.fudan.edu.cn}

\author{Xiaoyu Hong}
\affiliation{%
  \institution{Fudan University}
  \city{Shanghai}
  \country{China}}
\email{25113050033@m.fudan.edu.cn}

\author{Qingshan Li}
\affiliation{%
  \institution{Xidian University}
  \city{Xi'an}
  \country{China}}
\email{qshli@mail.xidian.edu.cn}

\author{Xin Peng}
\affiliation{%
  \institution{Fudan University}
  \city{Shanghai}
  \country{China}}
\email{pengxin@fudan.edu.cn}

\renewcommand{\shortauthors}{Trovato et al.}

\newcommand{\tool}{LogCopilot\xspace}

\begin{abstract}
Logs record the runtime behavior of software and are widely used in various tasks such as debugging, testing, and fault diagnosis.
With the increase in system size and complexity, log analysis has gradually become a challenging task.
Current industrial systems typically use log aggregation systems such as Grafana Loki and ELK to simplify the log collection and analysis process.
Engineers write queries using the DSL query language provided by these systems can complete a variety of log analysis tasks.
However, writing these queries is often time-consuming and labor-intensive, as it requires engineers to have a thorough understanding of the DSL syntax and the detailed information contained in the logs.
To address these challenges, this paper proposes \tool, an automated log aggregation analysis framework based on large language models (LLMs).
LogCopilot accepts natural language log analysis instructions and accomplishes automated log analysis through knowledge retrieval and tool calling.
\tool constructs a hierarchical knowledge base to represent and provide key knowledge in logs.
And it achieves automated log aggregation analysis by generating and executing LogQL queries.
The evaluation based on four log datasets confirm the effectiveness of \tool, which achieves an average accuracy of 76.8\% and outperforms baseline approaches.
Moreover, experiment results shows that \tool is effective in LogQL query generation.
\end{abstract}


\received{20 February 2007}
\received[revised]{12 March 2009}
\received[accepted]{5 June 2009}

\maketitle

\section{Introduction}

Software logs are semi-structured text that are continuously generated during the runtime of software systems.
Logs record the runtime behaviors of software systems, and are widely used in various tasks, such as debugging, testing, and fault diagnosis~\cite{log_survey}.
With the increasing scale of modern software systems and the application of new distributed architectures such as microservices, the collection, storage, and analysis of logs has gradually become challenging tasks~\cite{microsoft_loganalysis}.
Therefore, many log aggregation systems have emerged and are widely used in industrial software systems.

Widely used log aggregation systems such as Grafana Loki~\cite{loki}, ELK~\cite{elk}, Datadog~\cite{datadog}, etc., support the collection, standardization, and integration of log data from the entire software system.
Log aggregation systems typically provide DSL-based interfaces to simplify log analysis process, allowing engineers to query or aggregate logs by writing DSL.
For example, Grafana Loki implements the log query language LogQL~\cite{logql}.
Engineers can write LogQL queries to query specific logs or calculate log-related metrics based on the predefined operators or functions such as filtering and grouping.

However, writing the DSL queries also places high professional requirements on engineers.
Engineers need to understand the syntax of DSLs and need to understand the details of logs, such as format, level, variables, etc.
For example, the LogQL query shown in Figure~\ref{fig:logql} aims to count the number of times the remaining tickets for each train were queried in the past 30 minutes.
To write this query, engineers need to know the log format (the red segment in the figure) and the position of the train number variable in the log (the green segment in the figure).
In modern large-scale software systems, these requirements impose a significant learning cost on engineers.

Recently, researchers have also been exploring various technologies to assist with log analysis~\cite{qi2024logsay,locke2021logassist,seshagiri2024chatting,lognotion}.
LogAssist~\cite{locke2021logassist} and LogNotion~\cite{lognotion} help engineers quickly understand the various concepts and knowledge in logs by summarizing the logs.
However, they still rely on engineers to write the analysis statements themselves.
LogSay~\cite{qi2024logsay} achieves natural language-driven log numerical analysis by training a retriever and an inferencer. 
LogQLLM~\cite{seshagiri2024chatting} achieves natural language-to-LogQL translation by fine-tuning the LLMs.
However, both of the above approaches require the construction of large-scale labeled datasets, which entails significant manual labor costs.
Moreover, these models often have generalization issues, and in the rapidly evolving modern software systems, the models need to be retrained frequently.

To address the above challenges, this paper proposes \tool, an automated log aggregation analysis framework based on large language models.
\tool accepts natural language log analysis instructions and accomplishes automated log aggregation analysis through LLM-driven knowledge retrieval and tool calling.
Specifically, \tool first performs log analysis and workflow recognition to identify key knowledge in the logs, including variables, templates, and workflows. 
Then, \tool utilizes LLMs to hierarchically summarize the knowledge and construct a hierarchical knowledge base. 
During the log analysis phase, \tool supports two tasks: knowledge base question-answering (Q\&A) and log aggregation analysis. 
Given a log analysis instruction, \tool first understands the intent, then retrieves relevant knowledge from the knowledge base as needed. 
For log aggregation analysis tasks, it generates LogQL queries to call Grafana Loki for log aggregation analysis. 
Finally, it generates an analysis report based on the knowledge, aggregation analysis results, etc.

We evaluate \tool on four log datasets from previous studies~\cite{qi2024logsay,seshagiri2024chatting,zhu2023loghub,zhang2022deeptralog}.
Build upon existing works~\cite{qi2024logsay,seshagiri2024chatting}, we manually constructed a natural language log analysis question-answering benchmark based on these four datasets.
The experimental results prove that \tool can effectively perform log analysis and outperforms baseline approaches.
When using GPT-4o, its average answer accuracy can achieve 76.8\%.
Furthermore, experiments show that \tool is effective in LogQL query generation.

In summary, the main contributions of this work are:

\begin{itemize}
    \item We introduce \tool, a novel automated log aggregation analysis framework based on large language models.
    \item We conduct a series of experiments to demonstrate the effectiveness of \tool in automated log aggregation analysis.
    \item The code and dataset used in this paper are publicly available in the replication package~\cite{LogCopilot}.
\end{itemize}

\section{Background}

\subsection{Log Terminology}

Software logs, which are generated from various logging statements in the source code and are collected in an interleaving manner during the software execution, play a crucial role in the maintenance of industrial software systems. Logs record the system's runtime information and can be used to perform log aggregation analysis, extracting useful insights from massive logs.

\begin{figure}[t]
	\centering
	\includegraphics[width=0.6\linewidth ]{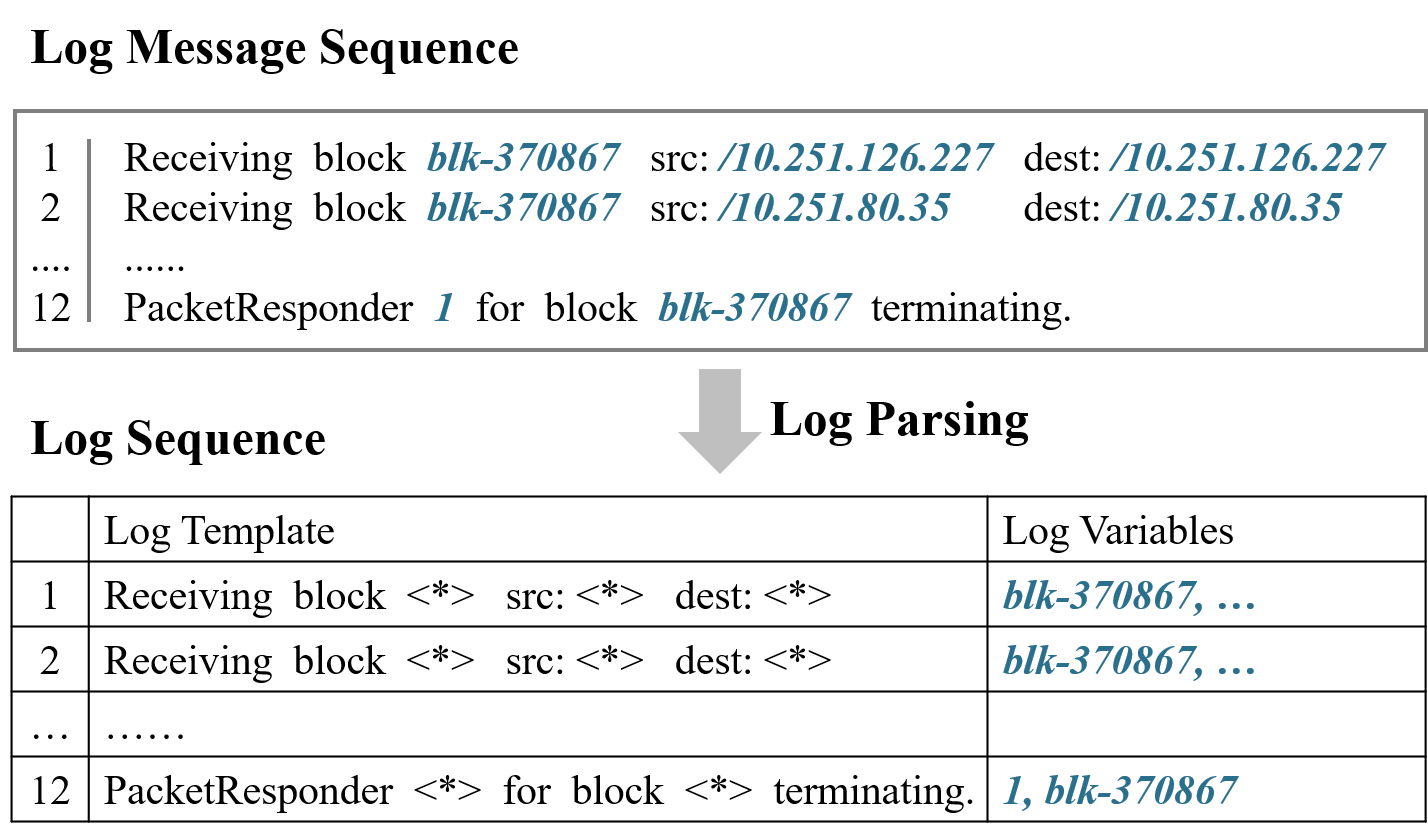}
	\caption{Example of the Log Variable, Log Template, and Log Sequence}
        \Description{An HDFS log example showing the relationship among raw log messages, log templates, log variables, and log sequences. The upper part lists raw log messages that
  share the same block identifier, with dynamic values such as block IDs and network addresses highlighted. The lower part shows how these messages are parsed into templates with
  placeholders and associated variables, and how the templates form a log sequence for the same task.}
	\label{fig:logs}
    \vspace{-3mm}
\end{figure}

Figure~\ref{fig:logs} shows an example of logs from a public dataset of HDFS~\cite{zhu2023loghub}. Each line in the upper part of Figure~\ref{fig:logs} is a log message, which is a raw semi-structured sentence. The log message consists of two elements: 1) log template: the fixed text written by the developers to describe a system event; 2) log variables: the values of the system attributes which carry dynamic runtime information, such as URL, file name, IP address, etc. The goal of log parsing is to distinguish the static and dynamic components from a log message. The output of a log parser is a structured log message, containing a log template and the key variables.

A series of log messages form a log message sequence. A log sequence is a series of log templates that record an execution flow of a specific task, which share the same task ID. The lower part of Figure~\ref{fig:logs} shows an example of log sequence with the task ID \textit{blk-370867}.

\subsection{Grafana Loki and LogQL}

To supprt efficient log analysis and monitoring in large-scale systems, log aggregation and management systems have become a significant infrastructure for efficient monitoring, debugging, and analytics. 
In recent years, Grafana Loki~\cite{loki}, an open-source log aggregation system, has been developed inspired by Prometheus~\cite{prometheus}. It offers a scalable and cost-effective solution for managing structured logs and is designed to be easy to operate. Instead of indexing the contents of the logs, but rather than a set of labels for each log stream. Inspired by Prometheus, Grafana Loki also provides a multi-dimensional log data model, an easy-to-use log query language, i.e., LogQL~\cite{logql}.


\begin{figure}[t]
	\centering
	\includegraphics[width=0.6\linewidth ]{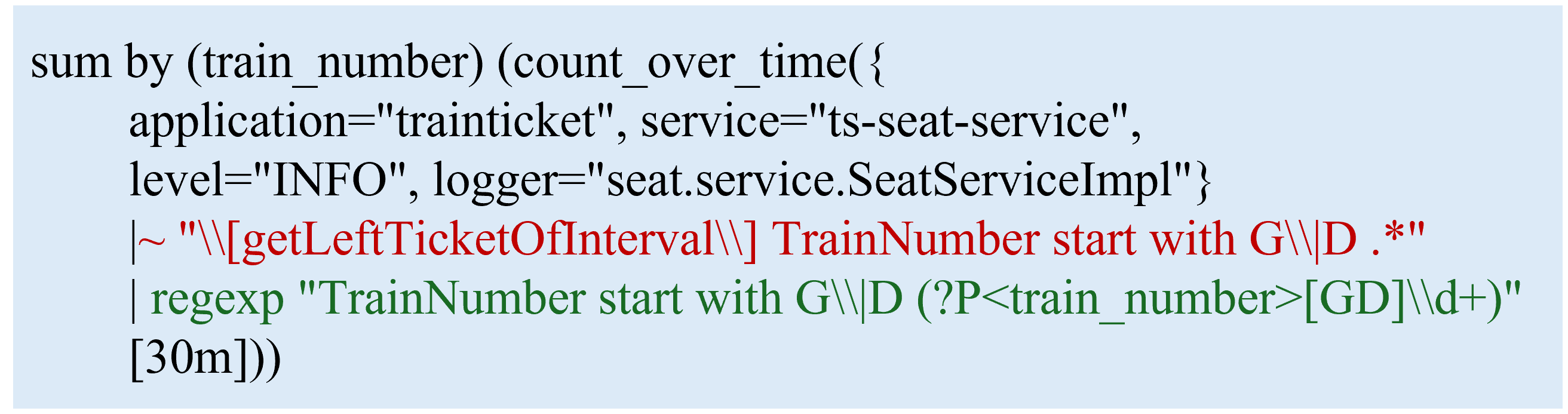}
	\caption{Example of the LogQL Query}
        \Description{A LogQL metric query for counting remaining-ticket query events by train number. The query selects TrainTicket service logs with specific stream labels, filters log
  lines for the getLeftTicketOfInterval operation, extracts the train number using a regular expression, counts matching logs over a 30-minute window, and sums the counts grouped
  by train number.}
	\label{fig:logql}
\end{figure}

To help engineers manipulate log data, log aggregation systems usually provide domain-specific languages (DSLs) offering interface to interact with log data. LogQL~\cite{logql} is Grafana Loki's log querying language, which is inspired by Prometheus' metric querying language called PromQL.
LogQL queries act as if they are a distributed \textit{grep} to aggregate log sources. There are two types of LogQL queries: log queries and metric queries.
Log queries return the contents of log lines. For example, \textit{\{container=``query-frontend''\} |= ``metrics.go''} selects log lines sourced from specific container and apply keyword filtering. 
Metric queries extend log queries by applying a function to log query results, creating metrics from logs. For example, \textit{count\_over\_time(\{job=``mysql''\}[5m])} counts all the log lines within the last five minutes for the MySQL job.
Figure~\ref{fig:logql} shows a typical LogQL query aiming at counting how many times the remaining ticket availability was queried for each train in the past 30 minutes. In this example, \textit{\{application=``trainticket'', service=``ts-seat-service'', level=``INFO'', logger=``seat.service.SeatServiceImpl''\}} is used to filter log streams based on the specified labels. The next two lines use regex expressions to filter the log lines and use regex capturing groups to extract labels from log lines respectively. Then the query uses a log range aggregation function to count the number of logs of each log stream during the specified time window \textit{[30m]}. Finally, an instant vector aggregation grouping by train number is applied to calculate the total occurrences for each train. Readers can refer to~\cite{logql} for more details of the LogQL.


\section{Approach Overview}

\begin{figure*}[t]
	\centering
	\includegraphics[width=\linewidth ]{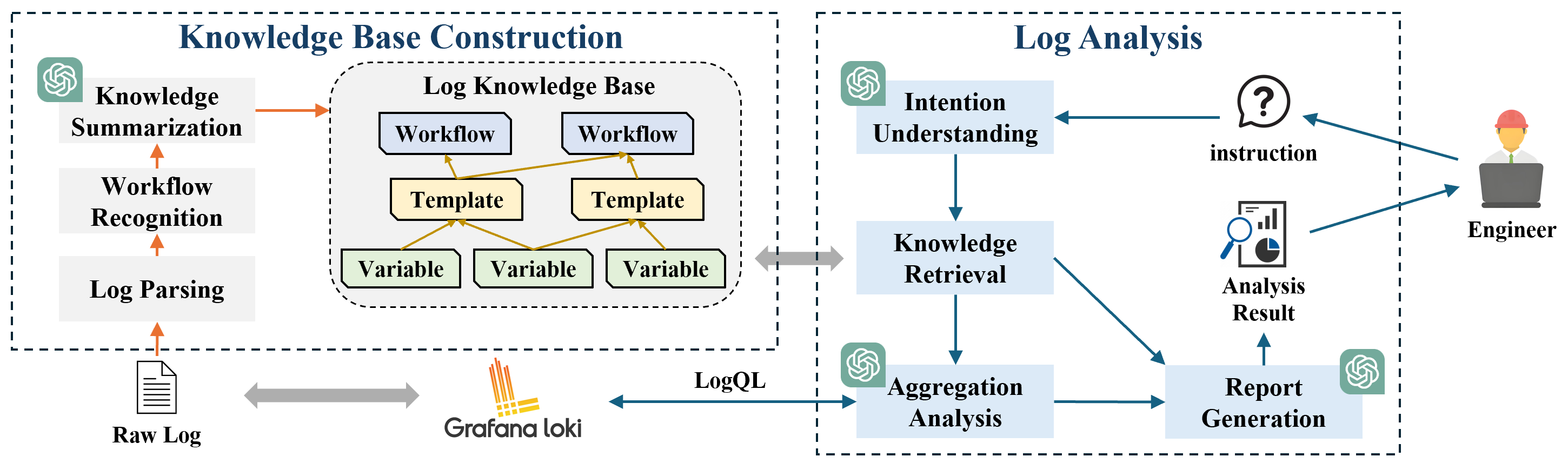}
	\caption{Overview of \tool}
        \Description{An overview pipeline of LogCopilot with two main parts. The offline knowledge base construction part processes raw logs through log parsing, workflow recognition,
  and GPT-based knowledge summarization to build a hierarchical log knowledge base containing workflows, templates, and variables. The online log analysis part takes an engineer's
  natural language instruction, performs intention understanding and knowledge retrieval, generates LogQL, queries Grafana Loki, performs aggregation analysis, and produces a
  final report.}
	\label{fig:overview}
\end{figure*}

\tool is a large language models-based log aggregation analysis framework.
It accepts natural language log analysis instructions and accomplishes automated log analysis through LLM-driven knowledge retrieval and tool calling. Figure~\ref{fig:overview} shows an overview of \tool, which consists of the following two phases.

\textbf{Phase-1: Offline Log Knowledge Base Construction}: In order to provide LLM with knowledge related to log analysis, \tool first constructs a log knowledge base by extracting and summarizing the variables, templates, and workflows in the raw logs.

\textbf{Phase-2: Online Log Analysis}: For a given log analysis instruction, \tool first retrieves the required knowledge from the knowledge base based on the intent of the instruction. 
And generates LogQL to call Grafana Loki to analyze the raw logs when required. 
Finally, the analysis report is generated based on the retrieved knowledge and the raw log analysis result.



\section{Knowledge Base Construction}
To support different granularity log aggregation analysis tasks, we focus on log knowledge in three dimensions: variables, templates, and workflow.
These three dimensions of knowledge contain information about log messages and log sequences.
And they have a hierarchical association relationship, which is that a variable may be used by different templates and a template may be included in different workflows.
To construct the knowledge base, we first perform log parsing to obtain log templates. Then we group the raw logs into log sequences and recognize workflows from them.
Finally, we summarize the semantic information of variables, templates, and workflows respectively using LLMs, and store them associatively as the final knowledge base.

\subsection{Log Parsing}

Following existing researches on log analysis~\cite{locke2021logassist,du2017deeplog,log_survey}, we obtain log templates through log parsing.
We adopt a state-of-the-art LLM-based log parsing approach, Lilac~\cite{jiang2024lilac}, which parses logs based on in-context learning and uses adaptive caching to avoid duplicate parsing.
We parse each raw log message into a log template and a list of variable values.
The log template contains the static part of a log, while variables in the log are replaced by placeholders, and each template is assigned an id (e.g., E1, E2).

\subsection{Workflow Recognition}
A sequence of log messages records the execution process of a certain task in the system.
Workflow is often used to represent a set of log sequences with similar behavior or corresponds to the same task, and has been widely used in existing researches~\cite{locke2021logassist,CloudSeer}.
Therefore, we recognize workflows from logs to represent knowledge related to log sequences.

Specifically, we follow the workflow construction method in LogAssist~\cite{locke2021logassist}.
We first group the raw logs based on grouping ids (e.g., task id, trace id, etc.) to obtain the log sequence. 
For logs without grouping ids, we group them by time window.
Then, we represent each log sequence as a sequence consisting of log templates. 
Finally, for repeating patterns (i.e., consecutive reoccurring templates or template combinations) in the sequence, we collapse them using the n-gram.
In this way, we represent each workflow as a concise sequence consisting of log templates.

\subsection{Knowledge Summarization} \label{subsec:knowledge_summarization}
The original variables, templates, and workflows lack high-level semantic information and are insufficient to support natural language-driven log analysis. 
Therefore, in this step, \tool uses LLM to summarizes the semantic information of variables, templates, and workflows. 
The summaries of each knowledge is represented as a paragraph of natural language and are saved as the knowledge base.
Because these three types of knowledge have a hierarchical relationship, we conduct a three-step hierarchical summarization process. 

\begin{figure}[t]
	\centering
	\includegraphics[width=0.75\linewidth ]{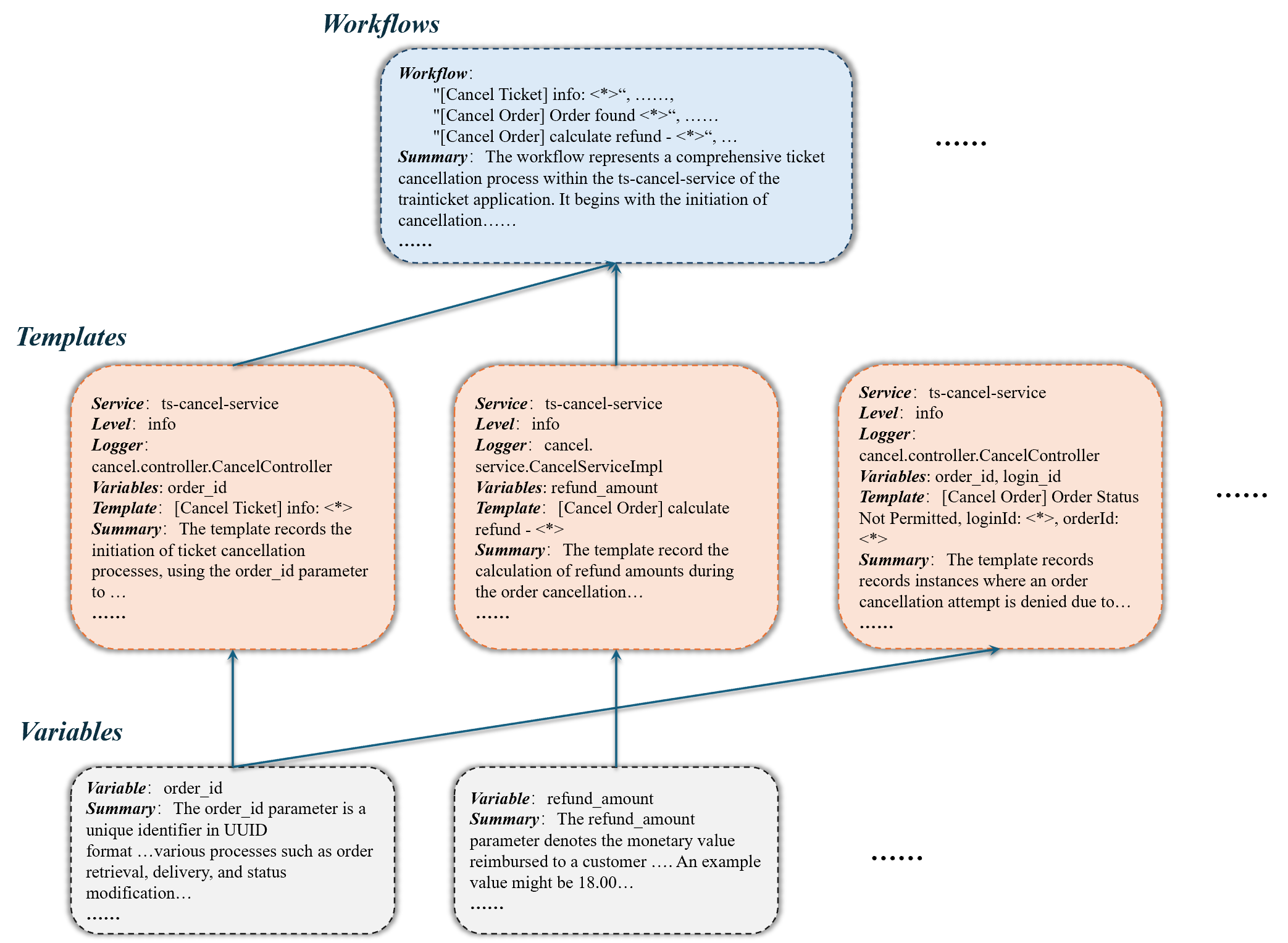}
	\caption{Example of the Log Knowledge Base}
        \Description{A hierarchical example of the LogCopilot log knowledge base. A workflow summary for a ticket cancellation process is shown at the top, several log template entries
  with service, level, logger, variables, template text, and summaries are shown in the middle, and variable entries such as order_id and refund_amount are shown at the bottom.
  Directed links indicate that variables are used by templates and templates are included in workflows.}
	\label{fig:knowledge base}
\end{figure}

\paragraph{Variable Summarization}
We use a two-stage method to recognize the semantics of variables and generate summaries for each variable.
First, we use LLMs to recognize the semantics of the variables contained in each template.
For each template, we find the workflows that contain that template and cluster those workflows using TF-IDF~\cite{bafna2016document} and k-means~\cite{macqueen1967some} (k is $3$ in this paper).
We then select the most representative workflows within each cluster, and for each workflow we randomly select one of its corresponding raw log sequences.
We include the template and these raw log sequences in a prompt, and prompt LLM to label the name of the variables in the template and generate a description for each variable.
A variable may appear in more than one log template.
Therefore, we feed variables with the same name along with their descriptions into the LLM in the second step.
And we prompt LLM to integrate these descriptions to generate the final summary of the variable.

\paragraph{Template Summarization}
Based on the obtained variable summaries, we summarize the templates.
We first use the same clustering method as in the variable summarization step to select representative log sequences for each template.
Then we include the template, the summaries of the variables in the template, and representative log sequences of this template in a designed prompt.
Finally, we prompt LLM to integrate this information to generate the summary of the template.

\paragraph{Workflow Summarization}
Similarly, we summarize the workflow based on the obtained template summaries.
We include the workflow, the summaries of the templates in the workflow, and several log sequences corresponding to this workflow in a designed prompt.
Then we prompt LLM to generate a summary of the workflow.

After the above steps, the final output is the three types of knowledge of the log in the system.
These three types of knowledge form a hierarchical relationship, as shown in the example in Figure~\ref{fig:knowledge base}.
It can be seen that each type of knowledge has a summary, and the information it contains, such as services and levels, can also be used for log analysis.

We use Elasticsearch~\cite{elasticsearch} to store the knowledge and the relationship between them.
And we use an embedding model (text-embedding-3-large~\cite{text-embedding-3-large} in this paper) to create embeddings for the summaries of variables, templates, and workflows, respectively.
All the summaries are embedded into 3072-dimensional vectors, and all the vectors are stored to support online retrieval.


\section{Log Analysis}
Based on the constructed log knowledge base and the log aggregation system Grafana Loki~\cite{loki}, \tool supports natural language instruction-driven automated log analysis.
\tool supports two types of log analysis tasks: log knowledge base Q\&A tasks, and raw log analysis tasks.
For a given log analysis instruction, \tool conducts log analysis in a four-step process: intention understanding, knowledge retrieval, log aggregation, and report generation.

\subsection{Intention Understanding}
Given a log analysis instruction, \tool first understands the intent of the instruction to clarify which log analysis task it corresponds to.
\tool designs a prompt that contains the description of the knowledge base, the functionality of Grafana Loki, and some representative log analysis tasks and their analysis process.
Then, we prompt LLM to identify which type of log analysis task the input instruction corresponds to.
For instructions that are difficult to understand, \tool returns the reasons to help users re-clarify the instructions.

\subsection{Knowledge Retrieval}
Based on the user intent, \tool retrieves the knowledge related to the log analysis task from the log knowledge base.
Since Grafana Loki mainly supports aggregated analysis of templates and variables, we retrieve different knowledge for the two types of tasks.

For the raw log analysis tasks, we retrieve template knowledge and variable knowledge. 
\tool uses the same embedding model as in the knowledge base construction step to generate the embedding of the log analysis instruction.
We retrieve the top-$K$ similar templates using the cosine similarity between the instruction and summary of the template.
Then, \tool retrieves the knowledge of the variables that are included in these templates from the knowledge base.

For the knowledge base Q\&A tasks, we retrieve all types of knowledge.
We generate the log analysis instruction embeddings in the same way.
However, for this type of task, we retrieve top-$K$ similar variables, templates, and workflows separately using cosine similarity.
Moreover, we follow the global search method in GraphRAG~\cite{edge2024local} to further score and rank the retrieved knowledge.
For each type of retrieved knowledge, we input it into LLM and let LLM determine which knowledge is helpful in answering the question, and generate key points and scores for each knowledge.
Finally, we rank each type of knowledge based on the generated scores as the final retrieval results.

\begin{figure}[t]
	\centering
	\includegraphics[width=0.5\linewidth ]{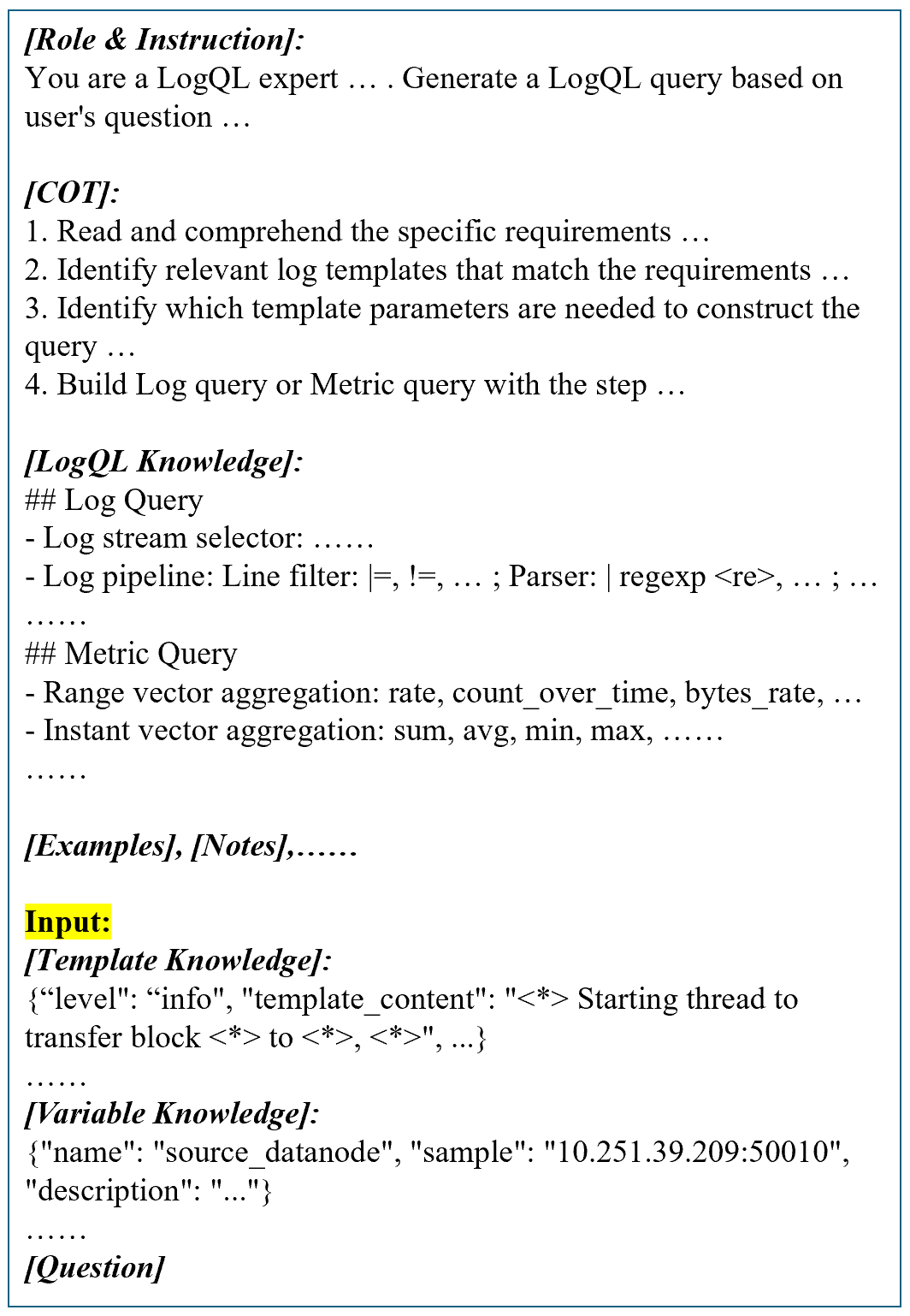}
	\caption{Prompt for LogQL Generation}
        \Description{A simplified prompt structure for LogQL generation. The prompt contains a role instruction for the model, chain-of-thought style steps for understanding the
  question and constructing the query, LogQL reference knowledge for log and metric queries, examples and notes, and input sections containing retrieved template knowledge,
  variable knowledge, and the user's question.}
	\label{fig:prompt}
\end{figure}

\subsection{Aggregation Analysis} \label{log_aggregation_analysis}
\tool implements aggregated analysis of raw logs by generating LogQL queries and calling Grafana Loki~\cite{loki}.
Based on the retrieved knowledge, \tool designs a prompt based on the paradigms of CoT and in-context learning.
Figure~\ref{fig:prompt} shows the structure of the prompt, the prompt has been simplified for clarity, while the complete version is available in our replication package.
Note that LogQL supports two aggregation modes, the first returns specific log messages, the second returns metrics calculated based on the raw logs.
Therefore, we describe these two aggregation modes in the prompt and include typical examples, thus enabling the model to generate LogQL queries as required.
And we include the retrieved template and variable knowledge in the prompt.

After generating LogQL query, \tool calls Grafana Loki to execute it.
The generated LogQL queries may have syntax or parameter errors that will cause execution failures, thus \tool designs a rectification process which is inspired by Reflexion \cite{shinn2023reflexion}.
When the execution fails, \tool adds the error message and the generated LogQL query to the prompt shown in Figure~\ref{fig:prompt} and prompts LLM to attempt repairing the LogQL query.
Then the repaired LogQL query will be executed again, we repeat the process at most three times to avoid getting into an infinite loop.
Finally, we will extract the log aggregation analysis results obtained from the execution as the final result.

\begin{figure}[t]
	\centering
	\includegraphics[width=0.6\linewidth ]{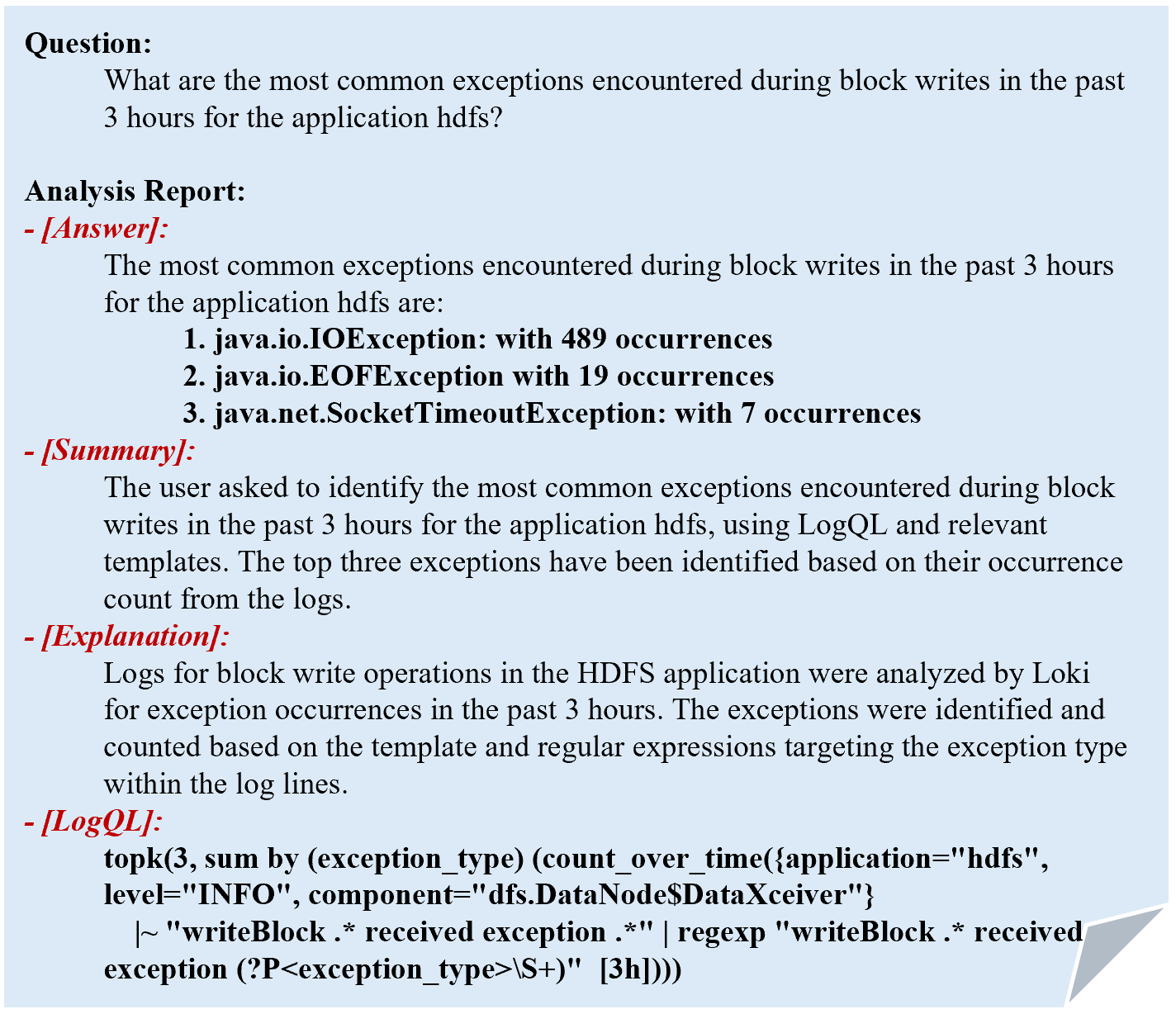}
	\caption{An Example of the Log Analysis Report}
        \Description{An example LogCopilot analysis report for an HDFS exception query. The figure shows the user's question, an answer listing the three most common exception types and
  their occurrence counts, a summary of the task, an explanation of how Loki logs were analyzed, and the generated LogQL query using topk, count_over_time, and regular-expression
  extraction of exception types.}
	\label{fig:report}
\end{figure}

\subsection{Report Generation}
Based on the retrieved knowledge and aggregation analysis results, \tool leverages LLMs to generate the final analysis report.
We designed a prompt to allow LLM to output a structured analysis report. 
For the knowledge base Q\&A tasks, we include all the retrieved knowledge in the prompt.
For the raw log analysis tasks, besides the retrieved knowledge, we also include the generated LogQL and the log aggregation analysis result in the prompt.

As shown in Figure~\ref{fig:report}, the analysis report is represented in natural language and consists of three fixed parts and one optional part.
Summary describes the log analysis task using a concise description.
Explanation describes and explains the entire analysis process in detail. 
LogQL shows the generated LogQL queries, and this part is only used in raw log analysis tasks.

\section{Evaluation}
We conduct a series of experimental studies to answer the following research questions.

\begin{itemize}
    \item \textbf{RQ1}: How effective is \tool in handling natural language log analysis tasks?
    \item \textbf{RQ2}: How much does knowledge summarization contribute to the effectiveness of \tool?
    \item \textbf{RQ3}: How does rectification affect the performance of LogQL queries generation?
    \item \textbf{RQ4}: How well does \tool perform in log knowledge retrieval?
\end{itemize}

\subsection{Experiment Setup}

\begin{table}
    \renewcommand{\arraystretch}{1.3}
    \setlength{\tabcolsep}{12pt}
    \centering
    \caption{Details of the Benchmark Dataset}
    \label{tab:datasets}
    \begin{tabular}{ccccc}
    \toprule
        System & Time Span & \# Messages & \# Templates & \# Tasks \\ \midrule
        HDFS & 38.7h & 11,167,740 & 46 & 100 \\
        OpenSSH & 682.4h & 638,947 & 38 & 100 \\
        OpenStack & 64.4h & 207,632 & 48 & 100 \\
        TrainTicket & 652.1h & 1,644,848 & 180 & 100 \\ \bottomrule
    \end{tabular}
\end{table}

\subsubsection{Datasets} 
To evaluate the effectiveness of \tool, we construct a new log analysis benchmark dataset based on logs from four systems: HDFS, OpenSSH, OpenStack, and TrainTicket~\cite{zhou2018fault}.
These four systems encompass various types, including microservices, cloud management platforms, big data platforms, etc., which can represent different log analysis scenarios.
On the basis of existing studies~\cite{qi2024logsay, seshagiri2024chatting}, we construct 400 log analysis tasks (100 tasks per system), each task consisting of a natural language question and the corresponding answer (including the analysis result and LogQL).
Detailed information of the benchmark dataset is listed in Table~\ref{tab:datasets}.

\paragraph{Log Source}


The HDFS, OpenSSH, and OpenStack logs are sourced from Loghub-2.0~\cite{jiang2024large}, which is a widely used benchmark dataset for log parsing tasks. 
We use all the log messages it provides.
TrainTicket is a medium-scale open source microservice system for railway ticketing and has been widely used in research on log analysis~\cite{yu2024deep, he2022empirical, jiang2023look}, trace analysis~\cite{chen2025tracezip, xie2025tracepicker, peng2022trace, guo2020graph, zhou2019latent}, and other AIOps tasks~\cite{yang2025hg, xv2025making, ren2024slim, yu2023nezha}.
The log data we use comes from the dataset in DeepTraLog~\cite{zhang2022deeptralog}, which collected 7,705,050 logs from TrainTicket by simulating user requests.
Our experiment uses 1,644,848 normal logs in this dataset.
Furthermore, these logs contain trace IDs, enabling us to achieve cross service log correlation and analysis in microservice systems.



\paragraph{Dataset Construction}

Although previous studies~\cite{qi2024logsay,seshagiri2024chatting} have constructed log analysis datasets based on HDFS, OpenSSH, and OpenStack, these datasets all suffer from some shortcomings.
LogSay~\cite{qi2024logsay} constructed a log numerical reasoning QA dataset with 1,972 question-answer pairs from three different system logs~\cite{zhu2023loghub}: HDFS, Spark, and OpenSSH.
However, the logs used by LogSay come from LogHub 1.0~\cite{zhu2023loghub}, where each system has only 2,000 log messages.
This limitation prevents the numerical reasoning tasks in LogSay from accurately reflecting the scenario of performing aggregate analysis on large volumes of logs in actual systems.
LogQLLM~\cite{seshagiri2024chatting} built a text-to-LogQL dataset containing 424 cases using logs from three systems (HDFS, OpenSSH, and OpenStack) from Loghub-2.0~\cite{zhu2023loghub}.
Although the dataset in LogQLLM contains a large number of log messages and a lengthy collection time span, it focuses solely on the text-to-LogQL task, thereby limiting its applicability to evaluating only a subset of log analysis scenarios.
Moreover, both datasets suffer from certain quality issues. 
First, the analysis tasks in them both contain significant duplication. 
For example, in LogSay, a single task may appear over ten times, differing only in the parameters under examination. 
Second, many problem descriptions are imprecise and exhibit considerable ambiguity.


Building upon these foundational efforts while addressing their limitations, we manually construct a log analysis benchmark dataset using the logs of four systems, as shown in Table~\ref{tab:datasets}. 
Three Ph.D. students have participated in the dataset construction process. First, two Ph.D. students analysis the LogSay and LogQLLM datasets and selected high-quality analysis tasks. Specifically, each student is asked to thoroughly understand and test each log analysis task, identifying cases where both task descriptions and answers are accurate and reasonable.
Finally, the two students merged their analysis results, discussed divergent cases, and deduplicated overlapping tasks. 
Through this process, we get 33 tasks from the LogSay dataset and 119 tasks from the LogQLLM dataset. 
These tasks are used as initial seeds, and the three PhD students are asked to instantiate them in the logs of different systems.
Subsequently, we reference log analysis tasks mentioned in the open-source community and existing research, and further supplemented new log analysis tasks based on the characteristics of each system's logs.
Finally, we let the three Ph.D. students cross-review the constructed tasks and invited a faculty member with industry experience to review all tasks.
For tasks with disputes, all participants discussed and modified them together.

The final benchmark comprises 100 curated log analysis tasks per system, yielding 400 executable tasks across four datasets, encompassing a wide range of use cases.
Each log analysis task in the benchmark includes a natural language question and its corresponding analysis result. 
For tasks involving log aggregation analysis, we also compose the corresponding LogQL query. 
Furthermore, we verify the correctness of all answers and LogQL queries through actual execution or manual analysis.

Additionally, we import all logs into Grafana Loki and package them as Docker images to ensure consistency across environments. 
Researchers can use these images to rapidly build test environments and evaluate log analysis tools.
We believe this benchmark dataset will facilitate future research in log analysis.




\subsubsection{Baselines} 

To evaluate the effectiveness, we compare our method with the following three baselines.

\begin{itemize}
    \item \textbf{Few-shot Learning}: It is a similar case retrieval-based approach. We first implement a Chain-of-Thought (CoT)~\cite{wei2022chain} prompting strategy for text-to-LogQL. Then for each test question, we retrieve top-$k$ semantically similar questions from historical question-answer pairs, injecting them as in-context demonstrations. The retrieval results of the generated LogQL queries are used as final answers.
    \item \textbf{LogSay~\cite{qi2024logsay}}: It implemented a multi-step ``\textit{Retriever-Reader}'' system for automated answering of log-related questions. It adopts a Siamese network~\cite{wang2022tolerance} and keyword-based filtering to retrieve relevant logs. QANet~\cite{yu2018fast} architecture is used to perform numerical reasoning, which yields the final answer. 
    \item \textbf{LogQLLM~\cite{seshagiri2024chatting}}: 
    This approach uses LoRA~\cite{hu2022lora} to fine-tune the generic LLMs to obtain the specialized text-to-LogQL model.
    In our experiment, we fine-tuned the GPT-4o model using the fine-tuning method and the 424 training cases provided in their paper.
    Similar to Few-shot Learning, the generated LogQL queries are executed to yield the final answers.
    
\end{itemize}

\subsubsection{Metrics} 

We adopt the following three metrics to measure the effectiveness of \tool and baseline approaches.

\begin{itemize}
    \item \textbf{Accuracy}: The percentage of correctly handled tasks out of all tasks. A task will be considered correctly handled only when the final answer output by the approach is the same as the ground-truth answer. Note that since the final answer is generated by LLMs and it may have multiple equivalent forms, we manually label the results and consider all equivalent forms as correct.
    \item \textbf{SyntaxAcc}: The percentage of syntactically correct LogQL queries out of all generated LogQL queries. A LogQL query is considered syntactically correct only if it passes Grafana Loki's syntax checking.
    \item \textbf{Recall}: To evaluate the effectiveness of \tool in log knowledge retrieval, we calculate Recall for each log analysis task in test dataset as follows:

    \begin{equation}
    Recall = \frac{\left| \mathcal{D} _{retrieved}\cap \mathcal{D} _{relevant} \right|}{\left| \mathcal{D} _{relevant} \right|} \label{eq:retrieval_recall}
    \end{equation}
    
    \(\mathcal{D} _{retrieved}\) is the retrieved documents from log knowledge base. \(\mathcal{D} _{relevant}\) is the ground-truth relevant documents for the task. This metric quantifies the ability of retrieval component to capture all knowledge prerequisites for accurate log analysis. Higher \(Recall\) indicates fewer information gaps in downstream analyzing.
\end{itemize}

\subsubsection{Implementation} 

We implement the prototype of \tool with Python 3.10 in approximately 2,210 lines of code. All experiments are conducted on a Ubuntu 22.04LTS server with 8 $\times$ Intel(R) Xeon(R) Platinum 8269CY 2.5GHz CPU and 32GB RAM.

We select two LLMs in our experiments, i.e., \textbf{GPT-4o}~\cite{openai2024gpt4o} and \textbf{GPT-4o-LogQL}. GPT-4o demonstrates impressive performance in reasoning and coding, especially for complex queries. GPT-4o-LogQL, the top-performing variant in LogQLLM's original experiments, is derived through fine-tuning of GPT-4o on the LogQLLM benchmark. Following standard practice, we partition the dataset into 80\% training and 20\% validation splits. We incorporate both generic (GPT-4o) and LogQL-specialized (GPT-4o-LogQL) LLMs to provide a more comprehensive view of the effectiveness of \tool and baseline approaches.
We set the temperature of all LLMs as zero to ensure the reproduction. 

\subsection{RQ1: Effectiveness of \tool}


\begin{table}
    \centering
    \caption{Comparision of Accuracy of \tool and baselines}
    \label{tab:overall_effectiveness}
    \renewcommand{\arraystretch}{1.3}
    \setlength{\tabcolsep}{12pt}
    \begin{tabular}{cccccc}
    \toprule
     \multirow{2.5}{*}{System} & \multirow{2.5}{*}{LogSay} & \multirow{2.5}{*}{Few-shot} & \multirow{2.5}{*}{LogQLLM} &  \multicolumn{2}{c}{\textbf{LogCopilot}} \\
    \cmidrule(){5-6}
     &  &  &  & \multicolumn{1}{c}{GPT-4o} & \multicolumn{1}{c}{GPT-4o-LogQL} \\
    \midrule
    HDFS & 0.100 & 0.180 & 0.060 & \textbf{0.850} & \textbf{0.850} \\
    OpenSSH & 0.070 & 0.480 & 0.220 & \textbf{0.730} & \textbf{0.700} \\
    OpenStack & 0.010 & 0.530 & 0.080 & \textbf{0.750} & \textbf{0.790} \\
    TrainTicket & 0.030 & 0.080 & 0.000 & \textbf{0.740} & \textbf{0.820} \\
    \midrule
    Average & 0.053 & 0.318 & 0.090 & \textbf{0.768} & \textbf{0.790} \\
    \bottomrule
    \end{tabular}%
     \vspace{-2mm}
\end{table}

In this experiment, we evaluate the effectiveness of \tool in handling natural language log analysis tasks. \tool employs both GPT-4o and GPT-4o-LogQL as the backbone model. We only adopt GPT-4o-LogQL as the backbone model of LogQLLM because its performance significantly exceeds that of the other two fine-tuned models. We adopt GPT-4o as the backbone model of Few-shot Learning baseline. The results of this evaluation are detailed in Table~\ref{tab:overall_effectiveness}, which provides a comparative analysis of five methods across four datasets in terms of Accuracy.

The experimental results shown in Table~\ref{tab:overall_effectiveness} demonstrate that \tool consistently outperforms all baseline methods, achieving the highest Accuracy across the HDFS, OpenSSH, OpenStack, and TrainTicket systems. Notably, both variants of \tool, using the generic model (GPT-4o) and the LogQL-specialized model (GPT-4o-LogQL), achieve strong average Accuracy exceeding 0.750. These results highlight the effectiveness of our approach in comprehending user intent and generating correct LogQL queries across diverse systems.

Among all the baseline methods, LogSay shows the poorest performance, with a maximum Accuracy of only 0.100 on the HDFS dataset and a minimum of just 0.010 on OpenStack. This is primarily due to the essence of LogSay. It integrates different neural networks in different phases, each requiring large amount of training data. In our experiments, we synthesize approximately 396 task instances using GPT-4o based on our benchmark to train LogSay, but the result remains unsatisfactory. LogSay fails to generalize effectively with limited resource, suggesting its low practicality for real-world natural language log analysis.

The Few-shot Learning baseline, which is based on GPT-4o with 5 historical examples, exhibits moderate performance on benchmark. It performs relatively well on OpenSSH and OpenStack. These results indicate that the guidance and knowledge provided by historical examples can help improve the performance of log analysis. However, its performance is still worse than \tool, since that few-shot prompting alone cannot cover all the knowledge required for complex tasks. Moreover, this method heavily relies on curated examples and lacks robustness when facing diverse systems.

LogQLLM refers to GPT-4o-LogQL and is expected to be specialized in LogQL syntax. However, its actual performance on our benchmark is poor, with an average Accuracy of only 0.090. It even underperforms the Few-shot Learning baseline, which learns not only log knowledge but also LogQL syntax from historical examples, across all the systems. This implies that only mastering the LogQL syntax without integrating semantic understanding or contextual knowledge significantly limits the effectiveness of the model.

In contrast, \tool using the same GPT-4o-LogQL achieves the highest average Accuracy of 0.790. This evidences the significance of log template knowledge in improving semantic understanding and generating correct LogQL queries, which may contain regex expressions and pattern filtering. To further investigate the impact of backbone models, we compare the two variants of \tool: one powered by GPT-4o and the other powered by GPT-4o-LogQL. The results show that \tool using GPT-4o-LogQL achieves a higher average Accuracy and outperforms \tool using GPT-4o on OpenStack and TrainTicket. This may be due to the better LogQL generation capability of GPT-4o-LogQL. However, GPT-4o-LogQL slightly underperforms GPT-4o on the OpenSSH dataset (0.030 absolute drop). We believe that this is caused by the randomness of LLMs even though we have set the temperature as zero.

\begin{figure}[t]
    \centering
    \raisebox{0.3\height}{
        \includegraphics[width=70pt]{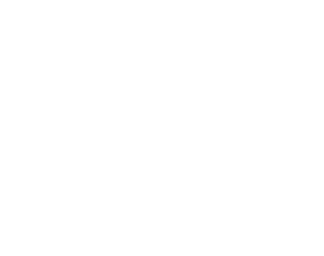}
    }
    \subfloat[HDFS] {
        \label{fig:venn:subfig:hdfs}
        \includegraphics[width=0.3\linewidth]{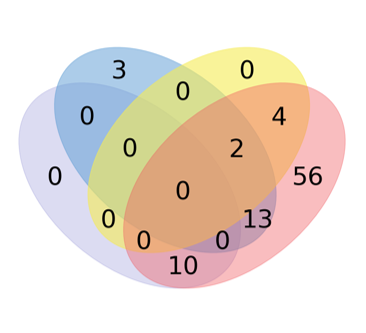}
    }
    \subfloat[OpenSSH] {
        \label{fig:venn:subfig:openssh}
        \includegraphics[width=0.3\linewidth]{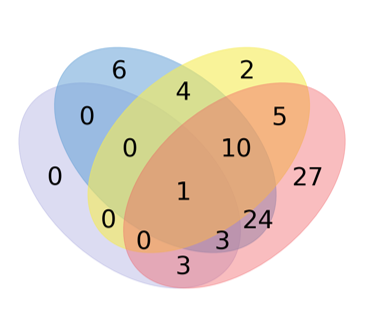}
    }
    \raisebox{0.6\height}{
        \includegraphics[width=70pt]{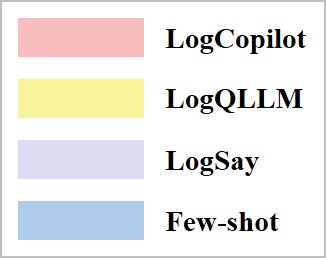}
    }
    
    \caption{Distribution of Questions Successfully Handled by \tool and Baselines}
    \Description{Two Venn diagrams comparing correctly answered questions across LogCopilot and three baselines on HDFS and OpenSSH. Each diagram contains four sets for LogCopilot,
  LogQLLM, LogSay, and Few-shot, with numbers indicating how many questions are correctly handled by each exact combination of approaches. The HDFS diagram shows many LogCopilot-
  only successes, including 56 questions solved only by LogCopilot, while the OpenSSH diagram also shows LogCopilot covering the largest number of unique successful cases.}
    \label{fig:venn}
\end{figure}

Figure~\ref{fig:venn} presents two Venn diagrams, demonstrating the unique questions successfully handled by each approach on HDFS and OpenSSH systems. These two diagrams highlight the complementarity and overlap among the approaches. Through Figure~\ref{fig:venn:subfig:hdfs}, we can find that \tool handles the most questions by understanding user intent and applying correct log aggregation analysis. Notably, \tool correctly answers 56 unique questions that fail in all baseline approaches. Only 2 questions in HDFS are covered by LogQLLM, Few-shot Learning, and \tool at the same time. This result underscores \tool's superior generalization and semantic understanding. Figure~\ref{fig:venn:subfig:openssh} indicates that \tool could solve 27 tasks that are beyond the capabilities of baseline approaches. Although the intersection of Few-shot Learning and \tool, with a size of 24, shows the basic capability of Few-shot Learning, 30 questions still remain unsolved by Few-shot Learning. The results emphasize limited capability of baseline approaches, suggesting that \tool can handle diverse cases that are typically challenging.

In conclusion, \tool leverages the semantic understanding of LLMs and retrieved log knowledge to perform log aggregation analysis, which is effective in answering log analysis questions and significantly outperforms baseline approaches. Compared to baseline approaches, \tool achieves higher Accuracy and covers more questions.

\subsection{RQ2: Contribution of knowledge summarization}

\begin{table}
    \renewcommand{\arraystretch}{1.3}
    \setlength{\tabcolsep}{12pt}
    \centering
    \caption{Accuracy of \tool without Knowledge Summarization}
    \label{tab:ablation_knowledge_summarization}
    \begin{tabular}{ccccc}
    \toprule
        Approach & HDFS & OpenSSH & OpenStack & TrainTicket \\ \midrule
        w/ KS & 0.850 & 0.730 & 0.750 & 0.740 \\ 
        w/o KS & 0.660\textsubscript{(\(\downarrow\)22.4\%)} & 0.640\textsubscript{(\(\downarrow\)12.3\%)} & 0.650\textsubscript{(\(\downarrow\)13.3\%)} & 0.670\textsubscript{(\(\downarrow\)9.5\%)} \\
        \bottomrule
    \end{tabular}
\end{table}

In Section~\ref{subsec:knowledge_summarization}, we have discussed knowledge summarization, which attaches high-level semantic information to variables, templates, and workflows. In this part, we conduct an ablation experiment to explore the contribution of log knowledge summaries generated by LLMs. To this end, we run \tool without knowledge summarization and compare the Accuracy with the full version. Instead of embedding the summaries of variables, templates, and workflows, \textbf{\tool w/o KS} creates embeddings based on the original content of variables, templates, and workflows.

Table~\ref{tab:ablation_knowledge_summarization} demonstrates the experimental results. Our observations indicate that the overall performance of \tool declines when knowledge summaries are lacking. Notably, HDFS suffers the most severe Accuracy degradation (22.4\%). This is because the semantics of individual templates in HDFS tend to be more low-level and system-centric, often reflecting internal operations such as data block management and node communication. As a result, there exists a semantic gap between these fine-grained log events and the high-level natural language questions. On the other three datasets, the Accuracy reduction ranges from 9.5\% to 13.3\%. This ablation study confirms the critical contribution of log knowledge summarization to \tool's effectiveness, particularly for systems with fine-grained log events.

\subsection{RQ3: Contribution of rectification}

\begin{table}
    \renewcommand{\arraystretch}{1.3}
    \setlength{\tabcolsep}{12pt}
    \centering
    \caption{Evaluation of the Contribution of Rectification}
    \vspace{-2mm}
    \label{tab:evaluation_rectification}
    \begin{tabular}{c c cccc}
        \toprule
        Metric& Approach & HDFS & OpenSSH & OpenStack & TrainTicket \\
        \midrule
        \multirow{2}{*}{SyntaxAcc}
        & w/o RE & 0.888 & 0.850 & 0.775 & 0.650 \\
        & w/ RE & 1.000\textsubscript{(\(\uparrow\)12.6\%)} & 0.938\textsubscript{(\(\uparrow\)10.4\%)} & 0.950\textsubscript{(\(\uparrow\)22.6\%)} & 0.850\textsubscript{(\(\uparrow\)30.8\%)} \\
        \midrule
        \multirow{2}{*}{Accuracy}
        & w/o RE & 0.800 & 0.650 & 0.613 & 0.575 \\
        & w/ RE & 0.850\textsubscript{(\(\uparrow\)6.3\%)} & 0.713\textsubscript{(\(\uparrow\)9.6\%)} & 0.750\textsubscript{(\(\uparrow\)22.4\%)} & 0.725\textsubscript{(\(\uparrow\)26.1\%)} \\ 
        \bottomrule
    \end{tabular}
\end{table}

In this part, we explore the effect of \tool's rectification mechanism mentioned in Section~\ref{log_aggregation_analysis}. It appends pending LogQL query with error message from Grafana Loki to context, and prompts LLM to repair the LogQL query by regeneration. Through controlled rectification experiments, we quantitatively evaluate \tool's performance improvement by comparing pre- and post-rectification results. We use \textbf{w/o RE} to indicate the results without rectification, and \textbf{w/ RE} to indicate the final results after at most three turns of rectification. Note that tasks not involving LogQL are not counted.

The experimental results are depicted in Table~\ref{tab:evaluation_rectification}. The results indicate that the rectification mechanism consistently improves SyntaxAcc and Accuracy across all datasets.
The SyntaxAcc on HDFS increases from 0.888 to a remarkable 1.000 after rectification. We observe the highest SyntaxAcc improvement (30.8\%) on TrainTicket. On the same system, the Accuracy improves from 0.575 to 0.725, corresponding to a relative gain of 26.1\%. Moreover, the Accuracy of all four systems exceeds 0.7.
These observations demonstrate the mechanism's capability to largely eliminate syntax errors in generated queries for complex log analysis tasks.

In terms of execution error type, one of the most common errors prior to rectification is the improper or missing use of aggregation functions. This frequently results in runtime errors such as ``maximum of series (500) reached for a single query", which occurs when LogQL query attempts to process too many time series without aggregating them appropriately. 

\subsection{RQ4: Recall of knowledge retrieval}

\begin{table}
    \renewcommand{\arraystretch}{1.3}
    \setlength{\tabcolsep}{6mm}
    \centering
    \caption{Recall of Log Knowledge Retrieval}
    \vspace{-2mm}
    \label{tab:knowledge_retrieval_recall}
    \begin{tabular}{cc}
        \toprule
        System & Average Recall \\
        \midrule
        HDFS & 0.978 \\
        OpenSSH & 0.958 \\
        OpenStack & 0.994 \\
        TrainTicket & 0.958 \\
        \bottomrule
    \end{tabular}
     \vspace{-2mm}
\end{table}

Recall of log knowledge retrieval is crucial to the effectiveness of \tool. Therefore, in this part, we evaluate the recall of \tool in retrieving log knowledge. We first label the ground truth of all the variables, templates and workflows that are essential in the LogQL query generation phase for each question. Then we calculate the recall rate for each question, based on Equation~\ref{eq:retrieval_recall}. To simplify the evaluation, we do not distinguish between the three types of knowledge documents in recall calculation. Instead, we treat all retrieved knowledge pieces as belonging to a unified set and compute recall based on whether the ground-truth items, regardless of their document type, appear in the top retrieved results. Finally, for each dataset, we average the results from all the questions to obtain the average recall.

Table~\ref{tab:knowledge_retrieval_recall} presents the average recall of \tool in knowledge retrieval. It can be seen that all of the average recall rates exceed 0.95, and \tool reaches the highest average recall of 0.994 on OpenStack. This indicates that \tool can effectively understand the intent of the question and retrieve the required variables, templates, and workflows.

\section{Discussion}
\subsection{Case Study}

\begin{figure}
	\centering
	\includegraphics[width=0.6\linewidth ]{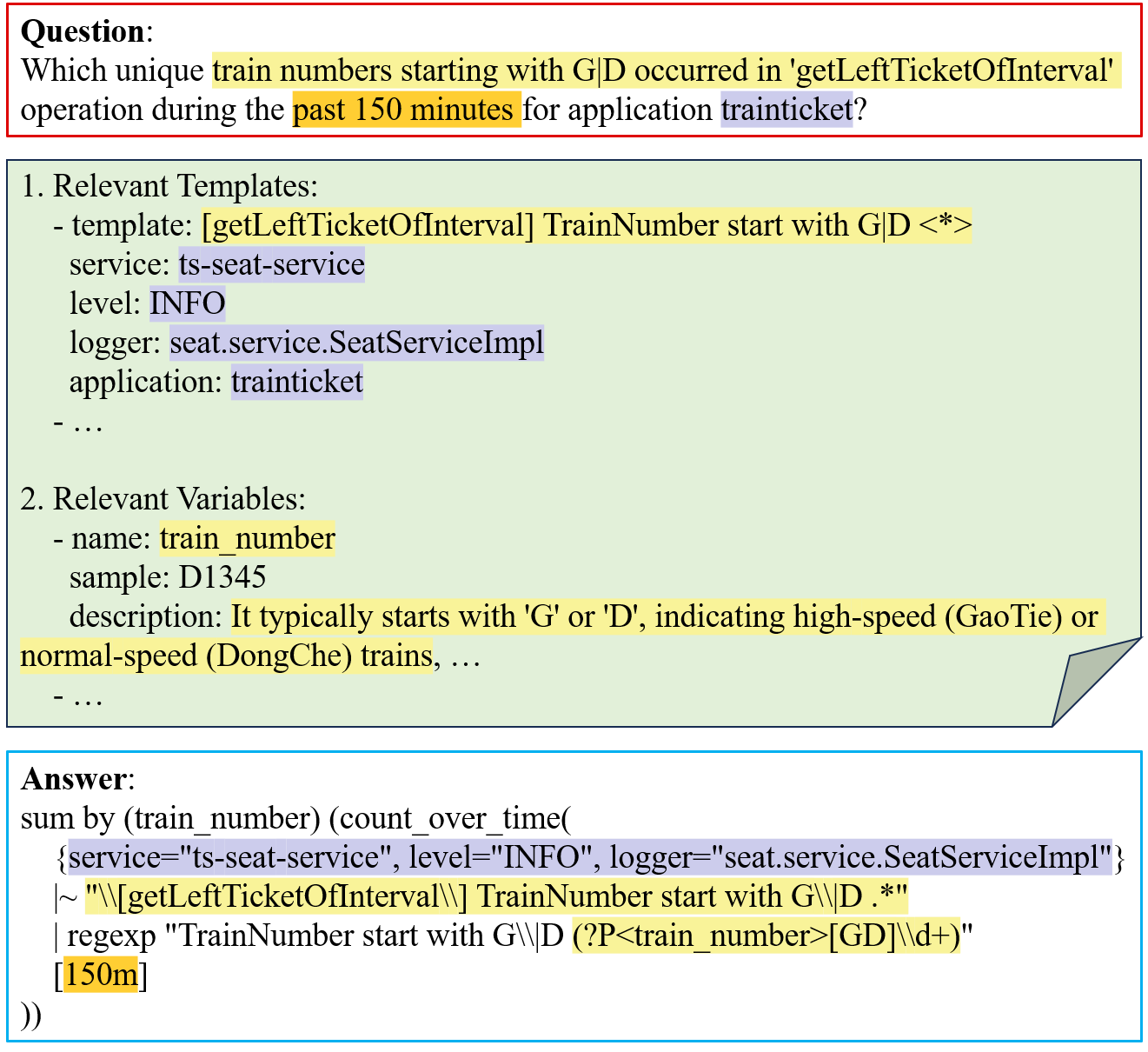}
	\caption{A Successful Case}
        \Description{A successful LogCopilot case for a TrainTicket query asking for unique train numbers starting with G or D in getLeftTicketOfInterval logs over the past 150 minutes. The figure shows the retrieved relevant template, the train_number variable and its description, and the generated LogQL query that filters the target service logs, extracts
  train_number with a regular expression, and aggregates counts by train number.}
	\label{fig:success_case}
\end{figure}

\begin{figure}
	\centering
	\includegraphics[width=0.6\linewidth ]{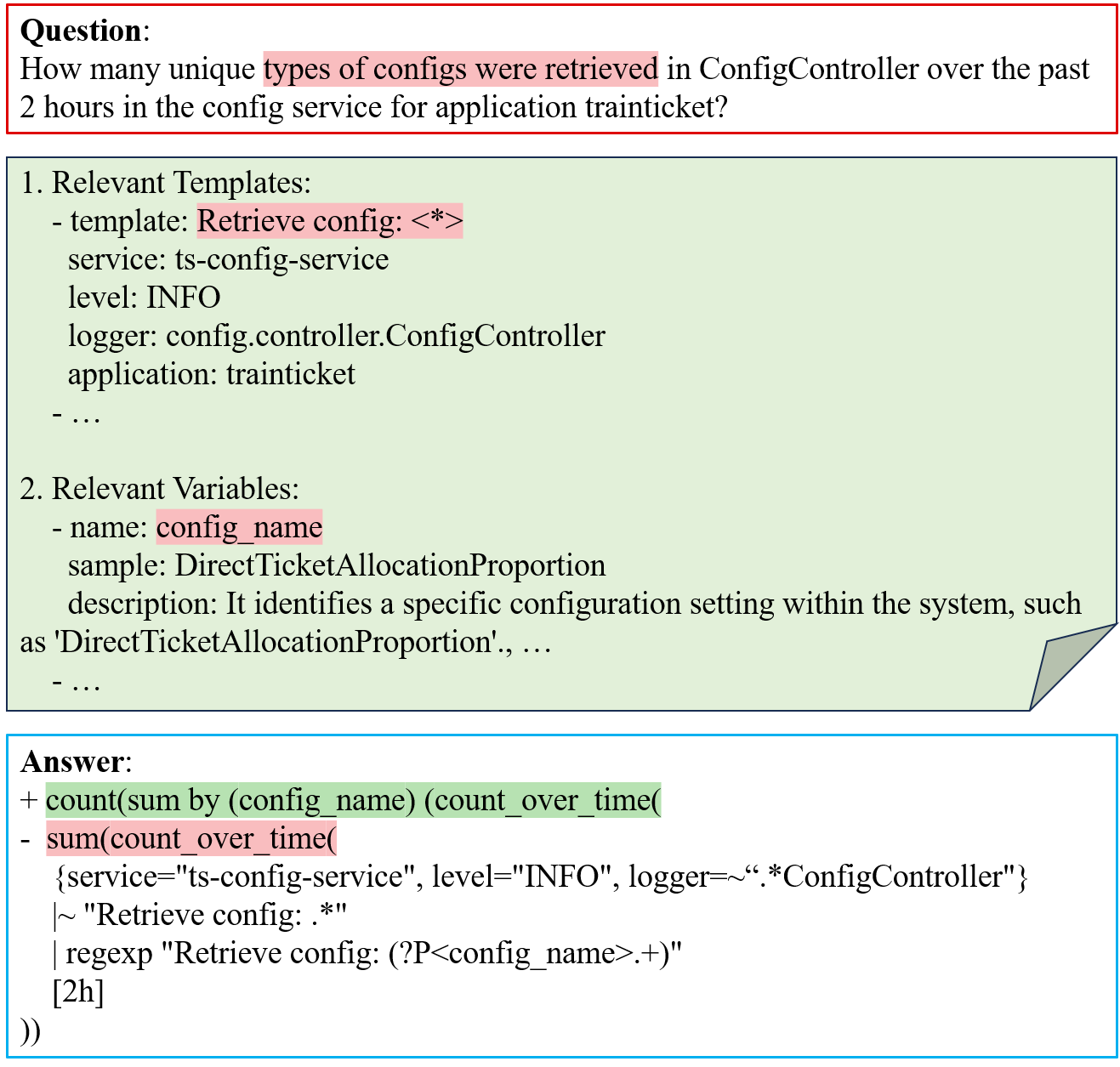}
	\caption{A Failed Case}
        \Description{A failed LogCopilot case for a TrainTicket query asking how many unique configuration types were retrieved by ConfigController over the past two hours. The figure
  shows that LogCopilot retrieves the relevant Retrieve config template and the config_name variable, but the generated LogQL query misses the intended aggregation structure for
  counting unique configuration names, leading to an incorrect final answer.}
	\label{fig:failure_case}
\end{figure}

To understand the capabilities and limitations of \tool, we further manually analyze a successful case and a failed case in this section. The two cases are shown in Figure~\ref{fig:success_case} and Figure~\ref{fig:failure_case}, where the green part is the relevant log knowledge retrieved by \tool. These cases are obtained based on \tool powered by GPT-4o-LogQL.

\subsubsection{Successful Case}

In Figure~\ref{fig:success_case}, we present a successful case where \tool understands user intent, performs correct log aggregation analysis using LogQL query, and generates answer matching the ground truth. In this case, the user wants to query the train numbers starting with ``G'' or ``D'' used in the ``getLeftTicketOfInterval'' operation. \tool performs intention understanding and generates LogQL query to aggregate logs. As shown in the yellow part of the figure, \tool first finds the semantically relevant log templates, e.g. ``[getLeftTicketOfInterval] TrainNumber start with \textless*\textgreater'', then finds the log variables corresponding to these log templates. As shown in the purple part of the figure, \tool selects the stream labels based on the retrieved log templates. Leveraging the above knowledge, \tool successfully generated the correct LogQL query.

In summary, \tool is able to effectively retrieve the knowledge related to the question and leverage the capabilities of LLMs to perform log aggregation analysis, yielding a natural language answer to the questioner.

\subsubsection{Failed Case}

In Figure~\ref{fig:failure_case}, we present a failed case where \tool fails to generate the correct LogQL query, yielding a wrong final answer. In this case, LogQL query generated by \tool is missing a grouping clause and a final verctor aggregation function. We highlight the ground-truth aggregation and the wrong aggregation by green and red respectively. However, we find that the relevant log templates and log variables have been retrieved by \tool, which is the red part of the knowledge in Figure~\ref{fig:failure_case}. It indicates that the retrieval method proposed by \tool is effective, but the randomness or limitations of the capabilities of LLMs can lead to incorrect results. On one hand, the LLMs often struggle to accurately capture the user's true intent due to the inherent vagueness of natural language. On the other hand, the LLMs may lack a deep understanding of the underlying data model, merely mimicking the syntax without fully grasping the semantics of the operations involved.

In summary, sometimes the LogQL generated by \tool is not precise. It is because the LLM does not really understand the underlying data model of Grafana Loki and the impacts of aggregation functions, even though the LLM has been fine-tuned to learn LogQL syntax. In our future work, we can use log aggregation tasks integrated with data model view to fine-tune LLMs, enhancing the LLMs' capabilities in both generating syntactically correct query and manipulating log streams as user's intent.

\subsection{Threats to Validity} 

\subsubsection{Internal Threat}

The threat to internal validity mainly lies in the implementation and configuration of baseline approaches. LogSay provides an open-source implementation but the code cannot be used on our benchmark directly. So we manually convert the format of samples in our benchmark and synthesize approximately 396 task instances as the training set using GPT-4o based on the transformed benchmark. LogQLLM has also provided publicly available fine-tuning source code and datasets. Due to the closed-source and proprietary of GPT-4o, we cannot access the GPT-4o fine-tuned by LogQLLM researchers. To alleviate the threat, we carefully follow the paper and use exactly the same script to fine-tune GPT-4o using OpenAI platform. There is no directly available open-source implementation of the few-shot learning-based approach. Therefore, we implement the approach by referring to the example of question answering using embeddings-based search provided in the OpenAI cookbook~\cite{openai2024cookbook}. To control the influences of the configurations of baseline approaches, we conduct experiments for all baseline approaches and select the best configuration for them.

\subsubsection{External Threat}

The external threats in our studies are mainly from the subject datasets and noises in labeling.

\textbf{Subject Datasets}: In this work, we have only evaluated \tool and three baseline approaches on four datasets. These log data comes from the distributed system (i.e., HDFS and OpenStack), server application (i.e., OpenSSH), and microservice system (i.e., TrainTicket). Although HDFS, OpenSSH, and OpenStack all come from real-world systems and contain millions of log messages, the number of subject systems is still limited and does not cover all the domains. Though TrainTicket is one of the largest open-source microservices systems and has been widely used in existing studies~\cite{soldani2022anomaly, zhang2022deeptralog, wu2021data, lee2023eadro}, there is still a gap between its scale and industrial large-scale microservices systems. Moreover, although we curate more diverse log analysis tasks than existing studies~\cite{qi2024logsay, seshagiri2024chatting}, these log analysis tasks cannot cover all situations. Therefore, the results of our experimental studies may not be generalized to larger or more complex datasets. In the future, we will evaluate our approach on more datasets.

\textbf{Noises in Labeling}: Our experiments are based on four log datasets that are widely used by related work~\cite{shan2024face, xu2024divlog, huo2023semparser, du2017deeplog}. The log analysis questions and corresponding answers are manually synthesized, as there are few well-constructed datasets for natural language log analysis. To evaluate the effectiveness of our approach, manual annotation is required to provide the ground-truth analysis result and LogQL queries for each task. Though this benchmark is manually constructed and carefully checked by PhD students and faculty, data noise may still be introduced during the manual labeling process. Although we believe the amount of noise is small, we will investigate the quantity issue in our future work.

\section{Related Work}
\textbf{Natural Language to DSL}: Domain-Specific Languages (DSLs) provide tailored syntactic expressions for particular problem spaces, or domains (e.g., SQL for database). In recent years, machine learning and deep learning techniques have been widely used in text-to-DSL tasks, such as text-to-SQL~\cite{ katsogiannis2023survey, kim2020natural} and text-to-GraphQL~\cite{ni2024knowledge}. These studies aim to automatically map natural language instructions to DSL, enabling users to interact with domain-specific tools or data through natural language.

Among these studies, text-to-SQL has emerged as a prominent area of study, due to the widespread adoption of SQL for structured data querying and database management tasks. With the advancement of Large Language Models (LLMs), a growing number of researchers have investigated the application of LLMs to text-to-SQL tasks \cite{li2024dawn, shi2024survey, liu2024survey, li2023can}. Pourreza and Rafiei~\cite{pourreza2023din} propose a decomposed in-context learning framework for text-to-SQL, achieving substantial improvements in execution accuracy. Pourreza et al.~\cite{pourreza2024chase} introduce a multi-path reasoning framework that enhances text-to-SQL performance by combining diverse candidate generation strategies. Ren et al.~\cite{ren2024purple} present PURPLE, a retrieval-augmented prompting framework that enhances LLM performance in NL2SQL tasks by selecting demonstrations containing relevant logical operator compositions. 

While extensive research has been conducted on text-to-SQL, much less attention has been paid to translating natural language into LogQL. To the best of our knowledge, there exists only one prior work that explicitly explores the task of text-to-LogQL. Seshagiri et al.~\cite{seshagiri2024chatting} fine-tune LLMs using manually curated text-to-LogQL datasets, gaining substantial improvements in LogQL generation.

\textbf{Log Analysis}: Logs, which record runtime information and key events, are playing an important role in system monitoring. Therefore, how to reasonably utilize log data has become a key area of research. Many efforts have been dedicated to log analysis, including log parsing, anomaly detection, root cause analysis, etc. 

Before performing log analysis, it is often essential to apply log parsing techniques to structure the raw log messages into a more analyzable form. Various automated log parsers have been proposed to extract templates or event types from unstructured logs, such as Drain~\cite{he2017drain}, UniParser~\cite{liu2022uniparser}, and SemParser~\cite{huo2023semparser}. Some prior studies try to assist practitioners in comprehending system workflows, to facilitate debugging and test case design. Locke et al.~\cite{locke2021logassist} propose LogAssist, which seeks to summarize logs into workflows to facilitate log analysis tasks. Logs can be utilized for anomaly detection. Du et al.~\cite{du2017deeplog} leverage a deep neural network model to detect anomalies based on recognized log sequences. Zhang et al.~\cite{zhang2019robust} implement LogRobust, leveraging an attention-based Bi-LSTM model to detect anomalies on unstable log data. Some prior studies try to help users answer log analysis questions. Qi et al.~\cite{qi2024logsay} propose LogSay, a multistep “Retriever-Reader” system for automated answering of log-related questions, adopt deep learning models to perform log numerical reasoning.

Approaches in previous research often relies on manually curated rules or are trained on limited datasets. Due to their limited generalizability, these methods often require substantial effort to adapt to new datasets or systems. Otherwise, their performance degrades significantly. LLMs have shown great capabilities in various domains, indicating their potential in log analysis. Some prior studies have explored the integration of LLMs with log analysis. Jiang et al.~\cite{jiang2024lilac} and Xu et al.~\cite{xu2024divlog} prompt LLMs to parse logs using few-shot learning, achieving promising performance. Shan et al.~\cite{shan2024face} propose a LLM-based strategy for localizing configuration errors through logs. Xu et al.~\cite{xu2024unilog} use LLMs to automatically decide where and what to log based on the in-context learning paradigm. In this paper, we propose a RAG-based log analysis framework to directly answer natural language questions, which is user-friendly compared to existing log analysis works.

\section{Conclusion}
In this paper, we propose \tool, an automated log aggregation analysis framework based on large language models. \tool accepts natural language log analysis instructions and accomplishes automated log analysis through LLM-driven knowledge retrieval and tool calling.
Specifically, \tool first performs log analysis and workflow recognition to identify key knowledge in the logs. Then, it hierarchically summarize the key knowledge and construct a hierarchical knowledge base. In the log analysis phase, \tool supports both knowledge based Q\&A and log aggregation analysis based on the hierarchical knowledge base.
To evaluate \tool, we construct a log aggregation analysis benchmark dataset based on log data from four different types of systems.
Experimental results demonstrate that \tool can perform log aggregation analysis effectively and outperform all the baseline approaches.

\section{Data Availability}

All the data and code of the work can be found in our replication package~\cite{LogCopilot}.

\bibliographystyle{ACM-Reference-Format}
\bibliography{reference}

\end{document}